\documentclass[pre,aps]{revtex4}
\usepackage{amssymb}
\usepackage{graphicx}

\begin{document}

\title{Universal low-temperature magnetic properties of the classical
and quantum dimerized ferromagnetic spin chain}
\author{D.~V.~Dmitriev}
\email{dmitriev@deom.chph.ras.ru}
\author{V.~Ya.~Krivnov}
\affiliation{Joint Institute of Chemical Physics of RAS, Kosygin
str. 4, 119334, Moscow, Russia.}
\date{}

\begin{abstract}
Low-temperature magnetic properties of both classical and quantum
dimerized ferromagnetic spin chains are studied. It is shown that at
low temperatures the classical dimerized model reduces to the
classical uniform model with the effective exchange integral
$J_{0}=J(1-\delta ^{2})$, where $\delta $ is the dimerization
parameter. The partition function and spin correlation function are
calculated by means of mapping to the continuum limit, which is
justified at low temperatures. In the continuum limit the
calculation of the partition function and spin correlation function
is reduced to the eigenvalue problem of quantum rotator in
gravitational field. Quantum model is studied using Dyson-Maleev
representation of the spin operators. It is shown that in the
long-wavelength limit the Hamiltonian of the quantum dimerized chain
reduces to that of the uniform ferromagnetic chain with the
effective exchange integral $J_{0}=J(1-\delta ^{2})$. This fact
implies that the known equivalence of the low-temperature magnetic
properties of classical and quantum ferromagnetic chains remains for
the dimerized chains. The considered model can be generalized to
include the next-neighbor antiferromagnetic interaction.
\end{abstract}

\maketitle

\section{Introduction}

The Paierls instability plays an important role in
quasi-one-dimensional materials. The Paierls metal-dielectric
transition originated from the coupling between electrons and
phonons occurs, as a rule, in organic solids \cite{book}. Such
transition can take place in the quantum spin chains coupled to
phonons as well (so-called the spin-Paierls transition (SP)). In
particular, the SP transition has been observed in the
antiferromagnetic spin chain $CuGeO_{3}$ \cite{MHase}. Currently
there is a growing interest to the quasi one-dimensional magnets
with ferromagnetic exchange interactions \cite{Landee,Delica,Sugano}
and the possibility of the Paierls instability in them is widely
discussed \cite{JSirker}. It is argued that the possible real system
where this instability takes place is monatomic chains of $Co$ on
the $Pt$ surface \cite{Gamb}. As it is proposed \cite{Gamb}, these
chains have the effectively ferromagnetic spin-spin interaction and
very weak elastic constants. Recently, another mechanism of the
Paierls instability in spin systems has been proposed
\cite{Ulrich,Horsch}. It is based on the coupling of the spins with
an electronic subsystem (spin-orbital mechanism). It is expected
that this mechanism is relevant to the transition metal oxide
$YVO_{3}$. The properties of this compound is described by the
spin-orbital model \cite{Khal}. The mean-field treatment of this
model leads to the 1D dimerized spin model with the ferromagnetic
sign of the interaction \cite{JSirker}.

Therefore, the study of the dimerized ferromagnetic (FM) chains is important
from both theoretical and experimental points of view. The spin chains with
the spin-phonon interaction are often described in the adiabatic
approximation which is valid if the phonon energy is smaller than the
Paierls gap. The Hamiltonian of this 1D spin model has the form%
\begin{equation}
H=H_{spin}+E_{elastic}  \label{H1}
\end{equation}%
where%
\begin{equation}
H_{spin}=J\sum_{n=1}^{N}(1-(-1)^{n}\delta )\mathbf{S}_{n}\cdot \mathbf{S}%
_{n+1}  \label{H2}
\end{equation}%
\begin{equation}
E_{elastic}=\frac{N\varkappa \delta ^{2}}{2}  \label{Eelastic}
\end{equation}%
where $\mathbf{S}$ is the spin operator, $J$ is the exchange
integral, $\delta $ is the dimerization parameter characterizing
lattice distortion, $\varkappa $ is the effective elastic constant.

There is essential difference between the antiferromagnetic (AF) and
ferromagnetic (FM) chains with respect to the coupling to the
lattice distortions. For the AF model ($J>0$) the ground state
energy of model (\ref{H2}) is $\sim -N\delta ^{4/3}$ \cite{Gross}
and the gain in this energy exceeds the loss in $E_{elastic}$ and
the SP transition takes place. Contrary to the AF model the ground
state of $H_{spin}$ at $J<0$ does not depend on $\delta $ and the
ground state of model (\ref{H1}) has the uniform lattice, $\delta
=0$. However, as was shown in Ref.\cite{JSirker} the thermal
fluctuations can activate the dimerization. It was shown in
Ref.\cite{JSirker} that the free energy of the FM chain at $T>0$ is
$-NT^{3/2}\delta ^{2}$ and the dimerized phase is stable at finite
temperature for small enough elastic constant $\varkappa $. Such
situation can occur in the system $Co$ chains on the $Pt$ surface
where the $Co$ atoms can be easily moved on the surface.

The dimerized FM chain can not been solved by the Bethe ansatz in
contrast with the uniform model with $\delta =0$. The thermodynamics
of this model has been studied in Ref.\cite{Herzog} using both
numerical TMRG simulations and the analytical modified spin-wave
theory \cite{Takahashi}. In particular, the phase diagram of the
model has been determined and the behavior of spin correlation
functions and the zero-field susceptibility are studied.

It is interesting to consider the influence of the external magnetic
field on the thermodynamics of the dimerized FM chain. The aim of
the present paper is to study the low-temperature magnetic
properties of model (\ref{H2}) independent of the dimerization
mechanism. Therefore, we will consider the model with the
Hamiltonian having a form
\begin{equation}
H=-J\sum (1-(-1)^{n}\delta )\mathbf{S}_{n}\cdot \mathbf{S}_{n+1}
-h\sum S_{n}^{z}  \label{H3}
\end{equation}
where $h$ is the dimensionless magnetic field and $J>0$.

Before we study this model it is instructive to note the remarkable fact
related to the uniform FM chain, $\delta =0$. It was claimed in Ref.\cite%
{sachdev} that the normalized magnetization $M=\left\langle
S^{z}\right\rangle /s$ of this model at $T\to 0$ is the universal
function of the scaling variable $g=s^{3}Jh/T^{2}$, i.e. the
universal function $M(g)$ is valid for any spin value $s$ and all
dependence on $s$ is captured in the scaling variable $g$. In
other words, this function is valid for both quantum and classical
ferromagnetic chains.

According to Ref.\cite{sachdev} the form of the function $M(g)$
can be determined by the computation of the magnetization of the
\textit{classical} FM chain in so-called scaling limit, when
$T\rightarrow 0$ and $h\to 0$ but the value of $g$ is finite. As
was shown in Ref.\cite{sachdev,Bacalis} the calculation of the
partition function of the classical model reduces to an eigenvalue
problem of a single quantum rotator. It allows to use the
efficient method to compute the function $M(g)$. As a result the
explicit expansions of this function in small and large $g$ were
obtained \cite{sachdev}.

We will show that the universal dependence of the magnetization
holds for dimerized model (\ref{H3}) as well. The corresponding
universal function coincides with that found in Ref.\cite{sachdev}
but the scaling variable $g$ is renormalized by a simple way to
include the dimerization parameter $\delta $, namely $g$ is
replaced by $\gamma =g(1-\delta ^{2})$.

The paper is organized as follows. In Section II we consider the
partition function of the dimerized classical ferromagnetic chain
in the magnetic field. We show that the field-dependent part of
the free energy coincides in the scaling limit with that for the
uniform model with the renormalized scaling parameter. In Section
III it is demonstrated that the spin-wave expansion of the
dimerized quantum ferromagnetic chain reproduces the large
$\gamma$ expansion of the magnetization of the classical model. In
Section IV the summary of the results is given and the
generalization to the model with the next-nearest neighbor
antiferromagnetic interaction is discussed.

\section{Classical dimerized spin chain in the scaling limit}

In this section we show that the low-temperature magnetic properties of the
classical dimerized spin model are reduced to that of the FM spin chain with
the renormalized exchange integral.

It is convenient to represent the Hamilton function of the considered
classical model (\ref{H3}) in a form%
\begin{equation}
H=-J_{1}\sum \vec{S}_{2i-1}\cdot \vec{S}_{2i}-J_{2}\sum \vec{S}_{2i}\cdot
\vec{S}_{2i+1}-\sum \vec{h}\cdot \vec{S}_{i}  \label{H4}
\end{equation}%
where $J_{1}=J(1+\delta )$, $J_{2}=J(1-\delta )$, and $\vec{S}_{i}$ are spin
vectors of the fixed length $s$ and the magnetic field is directed along the
Z axis: $\vec{h}=(0,0,h)$.

We represent spin vectors on odd and even sites as follows:
\begin{eqnarray}
\vec{S}_{2i-1} &=&s\vec{n}_{i}  \nonumber \\
\vec{S}_{2i} &=&s\vec{n}_{i}+s\vec{m}_{i}  \label{S-nm}
\end{eqnarray}%
where $\vec{n}_{i}$ are unit vectors and $\vec{m}_{i}$ are vector
differences between neighbor spins (we will assume $|\vec{m}_{i}|$ to be
small at low temperatures).

Then, the scalar products of spins on odd and even bonds become:
\begin{eqnarray}
\vec{S}_{2i-1}\cdot \vec{S}_{2i} &=&s^{2}-\frac{1}{2}(\vec{S}_{2i-1}-\vec{S}%
_{2i})^{2}=s^{2}-\frac{s^{2}}{2}\vec{m}_{i}^{2}  \nonumber \\
\vec{S}_{2i}\cdot \vec{S}_{2i+1} &=&s^{2}-\frac{s^{2}}{2}(\vec{n}%
_{i}^{\prime }-\vec{m}_{i})^{2}  \label{S-nm2}
\end{eqnarray}%
where we denoted
\begin{equation}
\vec{n}_{i}^{\prime }\equiv \vec{n}_{i+1}-\vec{n}_{i}  \label{np}
\end{equation}

After simple algebra the Hamilton function can be transformed to the form%
\begin{equation}
H=-\frac{h^{2}N}{2(J_{1}+J_{2})}+s^{2}\frac{J_{1}+J_{2}}{2}\sum \left( \vec{m%
}_{i}-\frac{sJ_{2}\vec{n}_{i}^{\prime }+\vec{h}}{s(J_{1}+J_{2})}\right) ^{2}+%
\frac{s^{2}J_{1}J_{2}}{2(J_{1}+J_{2})}\sum \vec{n}_{i}^{\prime 2}-2s\sum
\vec{h}\cdot \vec{n}_{i}  \label{Hnm}
\end{equation}

Then, the partition function of the system reads:
\begin{equation}
Z=\int\ldots\int\prod\limits_{i=1}^{N/2}d\vec{m}_{i}d\vec{n}_{i}\exp
\left( -\frac{H\left\{ \vec{m}_{i},\vec{n}_{i}\right\} }{T}\right)
\label{z1}
\end{equation}

We stress that up to here we did not do any assumption and
Eq.(\ref{Hnm}) and Eq.(\ref{z1}) are the exact expressions. Now we
assume that for low temperature $T\ll (J_{1}+J_{2})$ all vectors
$\vec{m}_{i}$ are small and directed in plane perpendicular to the
corresponding vectors $\vec{n}_{i}$. Then, we can integrate over
vectors $\vec{m}_{i}$ in the infinite limits. That gives
\begin{equation}
Z=e^{\frac{h^{2}N}{2T(J_{1}+J_{2})}}\left( \frac{2\pi
T}{s^{2}(J_{1}+J_{2})}\right) ^{N/2}
\int\ldots\int\prod\limits_{i=1}^{N/2}d\vec{n}_{i}\exp \left(
\frac{s^{2}J_{1}J_{2}}{T(J_{1}+J_{2})}\sum \left( \vec{n}_{i}\cdot
\vec{n}_{i+1}-1\right) +\frac{2s}{T}\sum \vec{h}\cdot
\vec{n}_{i}\right)  \label{z2}
\end{equation}

The first factor in Eq.(\ref{z2}) gives a constant contribution to
the magnetic susceptibility $\sim 1/J$. As will be shown below, in
the low-temperature limit the main contribution is given by the
integral in Eq.(\ref{z2}) and it is much higher ($\sim J/T^{2}$).
Therefore, we neglect the first factor in Eq.(\ref{z2}). The
second factor does not influence on the magnetic properties of the
system, and will be omitted. Thus, we reduced the partition
function of the dimerized chain to that of the uniform
ferromagnetic chain with the effective exchange integral
\begin{equation}
J_{0}=\frac{2J_{1}J_{2}}{J_{1}+J_{2}}=J(1-\delta ^{2})  \label{J0}
\end{equation}

The partition function of the classical FM chain in the low-temperature
limit was calculated in Ref.\cite{sachdev} by taking the continuum limit of
the model. We will follow this method. Partition function (\ref{z2}) in the
continuum approximation takes the form:
\begin{equation}
Z\varpropto \int D\left[ \vec{n}(x)\right] \exp \left( -\int_{0}^{L}\frac{dx%
}{2a}\left[ \frac{s^{2}J_{0}a^{2}}{T}\left( \frac{d\vec{n}}{dx}\right) ^{2}-%
\frac{2hs}{T}n_{z}\right] \right)  \label{z3}
\end{equation}%
where $L=Na$ and we notice that the distance between neighbor vectors $\vec{n%
}_{i}$ and $\vec{n}_{i+1}$ is two lattice spaces $2a$, so that the vector $%
\vec{n}_{i}$ corresponds to the vector field $\vec{n}(x)$ at the point $%
x=2ia $ in the continuum limit.

It is useful to transform Eq.(\ref{z3}) to dimensionless
variables. We rescale the spatial coordinate $x=ys^{2}aJ_{0}/T$
and obtain
\begin{equation}
Z\varpropto \int D\left[ \vec{n}(y)\right] \exp \left(
-\int_{0}^{\lambda }dy \left[ \frac{1}{2}\left(
\frac{d\vec{n}}{dy}\right) ^{2}-\gamma n_{z}\right] \right)
\label{z4}
\end{equation}%
where $\lambda =LT/as^{2}J_{0}$ is the scaled system length and
\begin{equation}
\gamma =\frac{s^{3}J_{0}h}{T^{2}}=g(1-\delta ^{2})  \label{gamma}
\end{equation}

Here $\gamma $ and $g$ are the scaling variables of the dimerized and
uniform models.

To calculate the partition function we utilize the well-known equivalence of
the $n$-dimensional statistical field theory with the ($n-1$)-dimensional
quantum field theory. The transition amplitude (or propagator) of a particle
located initially at $\vec{n}(0)=\vec{n}_{0}$, and finally at $\vec{n}%
(\lambda )=\vec{n}_{\lambda }$ takes the form of a path integral%
\begin{equation}
\left\langle \vec{n}_{\lambda }\right\vert e^{-\lambda \hat{H}}\left\vert
\vec{n}_{0}\right\rangle \propto \int_{\vec{n}_{0}}^{\vec{n}_{\lambda }}D[%
\vec{n}(y)]\exp \left\{ \int_{0}^{\lambda }L(\vec{n}^{\prime },\vec{n}%
)dy\right\}  \label{propag}
\end{equation}%
where $\hat{H}(\vec{n})$ is the Hamiltonian operator obtained by
quantization of the Hamilton function $H$ corresponding to the Lagrangian $L$%
.

Then, imposing the periodic boundary conditions $\vec{n}_{\lambda }=\vec{n}%
_{0}=\vec{n}$ and integrating over $\vec{n}$, we represent the partition
function (\ref{z4}) in a form%
\begin{equation}
Z\propto \int d\vec{n}\left\langle \vec{n}\right\vert e^{-\lambda \hat{H}%
}\left\vert \vec{n}\right\rangle  \label{Zpath}
\end{equation}

The quantum Hamiltonian has the form \cite{sachdev}:%
\begin{equation}
\hat{H}=\frac{\hat{l}^{2}}{2}-\gamma n_{z}  \label{Hquantum}
\end{equation}%
where $\hat{l}$ is an angular momentum operator. Hamiltonian (\ref{Hquantum}%
) describes the quantum rotator in the field $\gamma n_{z}$ and coincides
with the Hamiltonian for the uniform model \cite{sachdev} with $g$ replaced
by $\gamma $.

The corresponding Schr\"{o}dinger equation in the spherical coordinates has
the form:%
\begin{equation}
\left( -\frac{1}{2}\frac{d^{2}}{d\theta ^{2}}-\frac{\cot \theta }{2}\frac{d}{%
d\theta }-\frac{m^{2}}{2\sin ^{2}\theta }-\gamma \cos \theta \right) \psi
_{nm}=\varepsilon _{nm}\psi _{nm}  \label{sch}
\end{equation}%
where we used an axial symmetry of the model and introduced the azimuthal
quantum number $m$.

The exponent of the operator $\hat{H}(\vec{n})$ can be represented
using the eigenvalues and the eigenfunctions of the Schr\"{o}dinger
equation as
follows:%
\begin{equation}
e^{-\lambda \hat{H}(\vec{n})}=\sum_{nm}\left\vert \psi _{nm}(\vec{n}%
)\right\rangle e^{-\lambda \varepsilon _{nm}}\left\langle \psi _{nm}(\vec{n}%
)\right\vert  \label{expH}
\end{equation}

Then, the partition function becomes%
\begin{equation}
Z\propto \int d\vec{n}\left\langle \vec{n}\right\vert e^{-\lambda \hat{H}%
}\left\vert \vec{n}\right\rangle =\sum_{nm}e^{-\lambda \varepsilon _{nm}}
\label{z}
\end{equation}

In the thermodynamic limit $\lambda \to \infty $ only the lowest
eigenvalue $\varepsilon _{00}$ (with $m=0$) gives contribution to
the
partition function,%
\begin{equation}
Z\to e^{-\lambda \varepsilon _{00}}  \label{z6}
\end{equation}

The field-dependent part of the free energy per site is determined by the
ground state energy of Eq.(\ref{sch})%
\begin{equation}
F=\frac{T^{2}\varepsilon _{00}(\gamma )}{s^{2}J(1-\delta ^{2})}  \label{F}
\end{equation}

The normalized magnetization $M=\left\langle n^{z}\right\rangle $ is%
\begin{equation}
M=-\frac{\partial \varepsilon _{00}(\gamma )}{\partial \gamma }  \label{M}
\end{equation}

The solution of the Schr\"{o}dinger equation (\ref{sch}) has been
found analytically for small and large scaling parameter in
Ref.\cite{sachdev}. On the other hand, this equation can be solved
numerically for all values of $\gamma $ and the magnetization curve
can be found. It is shown on Fig.1. To demonstrate
$\delta$-dependence of the magnetization we represent it as a
function of the scaled magnetic field $g$ rather than $\gamma $. As
follows from Fig.1 the increase of the dimerization leads to the
decrease of the magnetization for all values of the magnetic field.

\begin{figure}[tbp]
\includegraphics[width=3in,angle=-90]{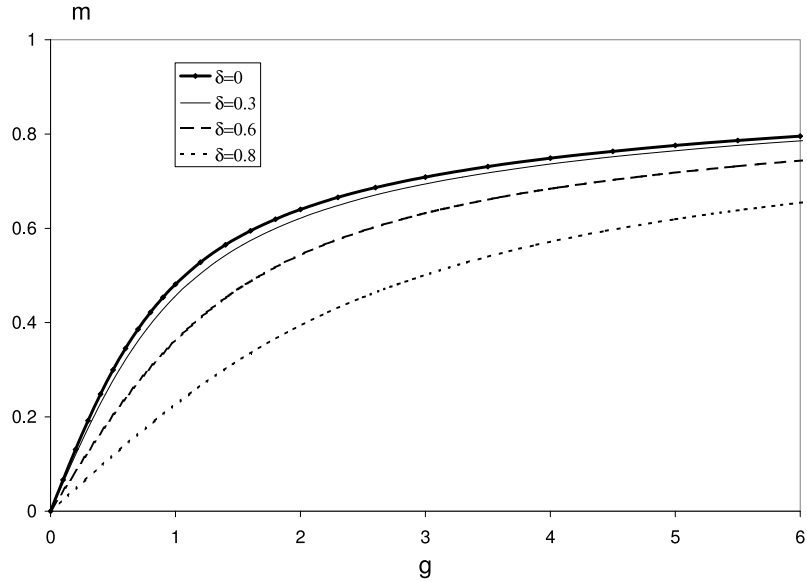}
\caption{Dependence of the normalized magnetization $M$ on the
scaled magnetic field $g=s^3Jh/T^2$ for several values of the
dimerization parameter $\delta$.}
\end{figure}

\subsection{Spin correlation functions}

The low-temperature magnetization of the dimerized FM chain
coincides with that for the uniform model and only the scaling
parameter $\gamma $ is renormalized. But the spin correlation
functions of the dimerized and uniform models are different
generally. Besides, the longitudinal and transverse correlators are
different for the non-zero magnetic field and we will consider both
types of correlators. At first we consider the correlators
$\left\langle S_{i}^{z}S_{i+2r}^{z}\right\rangle $ and $\left\langle
S_{i}^{x}S_{i+2r}^{x}\right\rangle =\left\langle
S_{i}^{y}S_{i+2r}^{y}\right\rangle $ for distances of even number of
lattice spacing $2r$. In this case the correlation functions do not
depend on $i$ and are defined by the reduced uniform model
(\ref{z2}):
\begin{equation}
\left\langle S_{i}^{\alpha }S_{i+2r}^{\alpha }\right\rangle
=s^{2}\left\langle n_{0}^{\alpha }n_{r}^{\alpha }\right\rangle
\end{equation}

The correlator $\left\langle n_{0}^{\alpha }n_{r}^{\alpha }\right\rangle $
can be expressed by the eigenvalues and the eigenfunctions of the Schr\"{o}%
dinger equation. We represent the brief derivation of these
expressions on the example of the correlator $\left\langle
n_{0}^{z}n_{r}^{z}\right\rangle $. The correlator $\left\langle
n_{0}^{z}n_{r}^{z}\right\rangle $ is by definition
\begin{equation}
\left\langle n_{0}^{z}n_{r}^{z}\right\rangle =\frac{1}{Z}\int D[\vec{n}%
(y)]n^{z}(0)n^{z}(\rho )\exp \left( \int_{0}^{\lambda }L[\vec{n}(y)]dy\right)
\end{equation}%
where
\begin{equation}
\rho =\frac{2Tr}{as^{2}J_{0}}  \label{rho}
\end{equation}

We divide the system on two parts $[0,\rho ]$ and $[\rho ,\lambda ]$. Then
the correlator is expressed as an integral of product of propagators:%
\begin{equation}
\left\langle n_{0}^{z}n_{r}^{z}\right\rangle =e^{\lambda \varepsilon
_{00}}\int n^{z}(0)d\vec{n}(0)\int n^{z}(\rho )d\vec{n}(\rho )I(0,\rho
)I(\rho ,\lambda )
\end{equation}

The propagators in the regions $[0,\rho ]$ and $[\rho ,\lambda ]$ are
calculated with the use of Eqs.(\ref{propag}) and (\ref{expH}):
\begin{equation}
I(0,\rho )=\int_{\vec{n}_{0}}^{\vec{n}_{\rho }}D[\vec{n}(y)]\exp \left(
\int_{0}^{\rho }L[\vec{n}(y)]dy\right) =\sum_{nm}\left\langle \vec{n}_{\rho
}|\psi _{nm}(\vec{n})\right\rangle e^{-\rho \varepsilon _{nm}}\left\langle
\psi _{nm}(\vec{n})|\vec{n}_{0}\right\rangle
\end{equation}%
and, similarly,%
\begin{equation}
I(\rho ,\lambda )=\sum_{nm}\left\langle \vec{n}_{\lambda }|\psi _{nm}(\vec{n}%
)\right\rangle e^{-(\lambda -\rho )\varepsilon _{nm}}\left\langle \psi _{nm}(%
\vec{n})|\vec{n}_{\rho }\right\rangle
\end{equation}%
where $\psi _{nm}$ and $\varepsilon _{nm}$ are the eigenfunctions and the
eigenvalues of the Schr\"{o}dinger equation (\ref{sch}).

Then, imposing the periodic boundary conditions $\vec{n}_{\lambda }=\vec{n}%
_{0}$ and using the definition $\hat{n}_{z}=\int d\vec{n}\left\vert \vec{n}%
\right\rangle n^{z}\left\langle \vec{n}\right\vert $ we obtain%
\begin{equation}
\left\langle n_{0}^{z}n_{r}^{z}\right\rangle =\sum_{nmn^{\prime }m^{\prime
}}e^{-\lambda (\varepsilon _{n^{\prime }m^{\prime }}-\varepsilon _{00})-\rho
(\varepsilon _{nm}-\varepsilon _{n^{\prime }m^{\prime }})}\left\vert
\left\langle \psi _{n^{\prime }m^{\prime }}|\hat{n}_{z}|\psi
_{nm}\right\rangle \right\vert ^{2}
\end{equation}

In the thermodynamic limit $\lambda \to \infty $ only the lowest
level in the region $[\rho ,\lambda ]$ survives, so that $n^{\prime
}=m^{\prime }=0$ and%
\begin{equation}
\left\langle n_{0}^{z}n_{r}^{z}\right\rangle =\sum_{nm}\left\vert
\left\langle \psi _{00}|\hat{n}_{z}|\psi _{nm}\right\rangle \right\vert
^{2}\exp \left( -\frac{T(\varepsilon _{nm}-\varepsilon _{00})}{%
s^{2}J(1-\delta ^{2})}2r\right)  \label{nzz}
\end{equation}

The expressions for other types of the correlation functions like $%
\left\langle n_{0}^{x}n_{r}^{x}\right\rangle $ has the same form as in Eq.(%
\ref{nzz}) with replacing $\hat{n}_{z}$ by $\hat{n}_{x}$:%
\begin{equation}
\left\langle n_{0}^{x}n_{r}^{x}\right\rangle =\sum_{nm}\left\vert
\left\langle \psi _{00}|\hat{n}_{x}|\psi _{nm}\right\rangle \right\vert
^{2}\exp \left( -\frac{T(\varepsilon _{nm}-\varepsilon _{00})}{%
s^{2}J(1-\delta ^{2})}2r\right)  \label{nxx}
\end{equation}

We expect that the universality in the long distance behavior of
the correlation functions holds, so we are interested in the
asymptotic of the correlation function $r\gg 1$. In this limit
only the lowest level(s) having non-zero matrix element makes
contribution to Eqs.(\ref{nzz}) and (\ref{nxx}). The operator
$n_{z}$ has non-zero expectation value over the ground state
$\left\langle \psi _{00}|\hat{n}_{z}|\psi _{00}\right\rangle $,
which is the normalized magnetization of the system $M$.
Therefore, the main contribution to sum (\ref{nzz}) is given by
$n=m=0$ term and equals $M^{2}$. The decaying correction to this
main term is given by the lowest excited state with the same
azimuthal number $m=0$, $\varepsilon _{10}$. Thus, the
long-distance asymptotic for the correlation function
$\left\langle n_{0}^{z}n_{r}^{z}\right\rangle $ is
\begin{equation}
\left\langle n_{0}^{z}n_{r}^{z}\right\rangle =M^{2}+\left\vert \left\langle
\psi _{00}|n_{z}|\psi _{10}\right\rangle \right\vert ^{2}\exp \left( -\frac{%
2r}{\xi _{\parallel }}\right)  \label{nzz3}
\end{equation}%
with the correlation length defined by the energy of the lowest excited
states as%
\begin{equation}
\xi _{\parallel }=\frac{s^{2}J(1-\delta ^{2})}{T\left( \varepsilon
_{10}-\varepsilon _{00}\right) }
\end{equation}

The operator $n_{x}$ changes the azimuthal number $m$, therefore the lowest
level for correlator $\left\langle n_{0}^{x}n_{r}^{x}\right\rangle $ is $%
\varepsilon _{01}$. This implies that the transverse correlation function
does not show the long range order and exponentially decays on large
distances:%
\begin{eqnarray}
\left\langle n_{0}^{x}n_{r}^{x}\right\rangle &=&\left\vert \left\langle \psi
_{00}|n_{x}|\psi _{01}\right\rangle \right\vert ^{2}\exp \left( -\frac{2r}{%
\xi _{\perp }}\right)  \nonumber \\
\xi _{\perp } &=&\frac{s^{2}J(1-\delta ^{2})}{T\left( \varepsilon
_{01}-\varepsilon _{00}\right) }  \label{nxx3}
\end{eqnarray}

The correlation lengths $\xi _{\parallel }$ and $\xi _{\perp }$
and the preexponential factors can be found analytically in the
limits $\gamma \to 0$ and $\gamma \to \infty $. At $\gamma =0$ the
Schr\"{o}dinger equation (\ref{sch}) reduces to the equation for
the operator of angular momentum with well-known spherical
eigenfunctions and the spectrum $l(l+1)/2$. So the correlation
functions for large $r$ are
\begin{equation}
\left\langle S_{i}^{z}S_{i+2r}^{z}\right\rangle =\left\langle
S_{i}^{x}S_{i+2r}^{x}\right\rangle =\frac{s^{2}}{3}\exp (-2r/\xi )
\label{Kg0}
\end{equation}%
where $\xi _{\parallel }=\xi _{\perp }=\xi $ and
\begin{equation}
\xi =\frac{s^{2}J(1-\delta ^{2})}{T}
\end{equation}

In the limit of high magnetic field ($\gamma \gg 1$) the lowest
eigenfunctions and the corresponding eigenvalues are
\begin{eqnarray}
\psi _{00} &=&2\gamma ^{1/4}e^{-\sqrt{\gamma }\theta ^{2}/2},\qquad
\varepsilon _{00}=-\gamma +\sqrt{\gamma }  \nonumber \\
\psi _{01} &=&\gamma ^{1/4}\theta \psi _{00},\qquad \varepsilon
_{01}=-\gamma +2\sqrt{\gamma }  \nonumber \\
\psi _{10} &=&(\sqrt{\gamma }\theta ^{2}-1)\psi _{00},\qquad \varepsilon
_{10}=-\gamma +3\sqrt{\gamma }
\end{eqnarray}

The correlation functions in the limit $\gamma \to \infty $ are%
\begin{eqnarray}
\left\langle S_{i}^{z}S_{i+2r}^{z}\right\rangle &=&s^{2}M^{2}+\frac{s^{2}}{%
4\gamma }\exp (-2r/\xi _{\parallel }),\qquad \xi _{\parallel }=\frac{1}{2}%
\sqrt{\frac{sJ(1-\delta ^{2})}{h}}  \nonumber \\
\left\langle S_{i}^{x}S_{i+2r}^{x}\right\rangle &=&\frac{s^{2}}{\sqrt{\gamma
}}\exp (-2r/\xi _{\perp }),\qquad \xi _{\perp }=\sqrt{\frac{sJ(1-\delta ^{2})%
}{h}}  \label{Kginf}
\end{eqnarray}

According to Eqs.(\ref{Kg0}) and (\ref{Kginf}) the correlation lengths are
changed from $\xi \sim 1/T$ at $\gamma =0$ to $\xi \sim h^{-1/2}$ for $%
\gamma \to \infty $. The crossover between two types of the behavior
of $\xi $ occurs at $\gamma \simeq 1$. The dependencies of
$\xi_\perp$ on $g$ for some values of $\delta $ are shown on Fig.2.

\begin{figure}[tbp]
\includegraphics[width=3in,angle=-90]{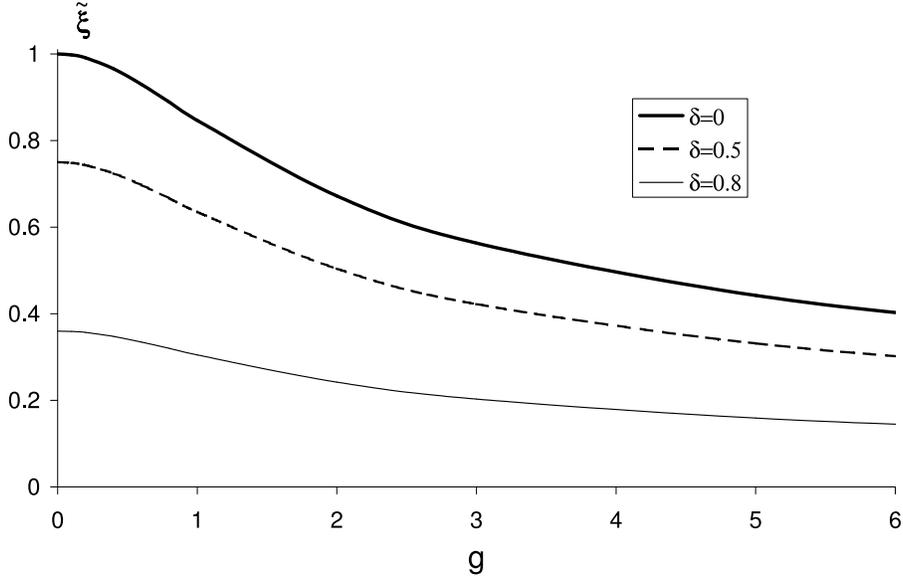}
\caption{Dependencies of $\tilde{\xi}=\xi_\perp T/s^2J$ on the
scaled magnetic field $g=s^3Jh/T^2$ for dimerization parameters
$\delta=0,0.5,0.8$.}
\end{figure}

Now we study the spin correlation function on `odd' distances. In
this case the correlation function $\left\langle S_{i}^{\alpha
}S_{i+2r+1}^{\alpha }\right\rangle $ is different for odd and even
$i$. Therefore, we distinguish two types of `odd' correlators:
$\left\langle S_{2j-1}^{\alpha }S_{2j+2r}^{\alpha }\right\rangle $
and $\left\langle S_{2j}^{\alpha }S_{2j+2r+1}^{\alpha
}\right\rangle $. The longitudinal correlators $\left\langle
S_{i}^{z}S_{i+2r+1}^{z}\right\rangle $ has non-zero asymptotic
$M^{2}$ at $r\to \infty $ and the calculation of the small
corrections caused by the dimerization to this value is not
important. On the contrary, the transverse correlator
$\left\langle S_{i}^{x}S_{i+2r+1}^{x}\right\rangle $ decays
exponentially and it is more interesting object for the
calculation of subtle effects like difference on odd and even
distances.

According to Eq.(\ref{S-nm}), the transverse correlator on odd distances is
represented as:
\begin{equation}
\left\langle S_{2j-1}^{x}S_{2j+2r}^{x}\right\rangle =s^{2}\left\langle
n_{0}^{x}(n_{r}^{x}+m_{r}^{x})\right\rangle
\end{equation}

The correlator $\left\langle n_{0}^{x}n_{r}^{x}\right\rangle $ was found
above in Eq.(\ref{nxx3}). For the correlator $\left\langle
n_{0}^{x}m_{r}^{x}\right\rangle $ we use the following identity:%
\begin{equation}
\int_{-\infty }^{\infty }xe^{-\alpha \left( x-y\right)
^{2}}dx=y\int_{-\infty }^{\infty }e^{-\alpha \left( x-y\right) ^{2}}dx
\end{equation}

Therefore, the integration over $\vec{m}_{r}$ in the multiple integral
\begin{equation}
\left\langle n_{0}^{x}m_{r}^{x}\right\rangle =\int\ldots\int\prod d\vec{m}_{i}d%
\vec{n}_{i}n_{0}^{x}m_{r}^{x}\exp \left( -\frac{H\left\{ \vec{m}_{i},\vec{n}%
_{i}\right\} }{T}\right)
\end{equation}%
can be transformed as
\begin{equation}
\int m_{r}^{x}\exp \left[ -\frac{J_{1}+J_{2}}{2T}\left( \vec{m}_{r}-\frac{%
sJ_{2}\vec{n}_{r}^{\prime }+\vec{h}}{s(J_{1}+J_{2})}\right) ^{2}\right] d%
\vec{m}_{r}=\frac{J_{2}n_{r}^{\prime x}}{J_{1}+J_{2}}\int \exp \left[ -\frac{%
J_{1}+J_{2}}{2T}\left( \vec{m}_{r}-\frac{sJ_{2}\vec{n}_{r}^{\prime }+\vec{h}%
}{s(J_{1}+J_{2})}\right) ^{2}\right] d\vec{m}_{r}
\end{equation}%
so that%
\begin{equation}
\left\langle n_{0}^{x}m_{r}^{x}\right\rangle =\frac{J_{2}}{J_{1}+J_{2}}%
\left\langle n_{0}^{x}n_{r}^{\prime x}\right\rangle
\end{equation}

Using the definition $n_{r}^{\prime x}=n_{r+1}^{x}-n_{r}^{x}$ and Eq.(\ref%
{nxx3}) we obtain
\begin{equation}
\left\langle S_{2j-1}^{x}S_{2j+2r}^{x}\right\rangle =s^{2}\left\vert
\left\langle \psi _{00}|n_{x}|\psi _{01}\right\rangle \right\vert
^{2}e^{-2r/\xi _{\perp }}\left( 1-\frac{J_{2}(1-e^{-2/\xi _{\perp }})}{%
J_{1}+J_{2}}\right)  \label{eq}
\end{equation}

Here we note that for $\xi _{\perp }\gg 1$, which is always assumed
for low temperature limit, the last factor in Eq.(\ref{eq}) can be
expanded and the
correlator takes the form%
\begin{equation}
\left\langle S_{2j-1}^{x}S_{2j+2r}^{x}\right\rangle =s^{2}\left\vert
\left\langle \psi _{00}|n_{x}|\psi _{01}\right\rangle \right\vert
^{2}e^{-(2r+1)/\xi _{\perp }}\left( 1-\frac{\delta }{\xi _{\perp }}\right)
\label{xx1}
\end{equation}

Similar, for the correlator $\left\langle
S_{2j}^{x}S_{2j+2r+1}^{x}\right\rangle $ we need merely to exchange $%
J_{2}\longleftrightarrow J_{1}$, which gives%
\begin{equation}
\left\langle S_{2j}^{x}S_{2j+2r+1}^{x}\right\rangle =s^{2}\left\vert
\left\langle \psi _{00}|n_{x}|\psi _{01}\right\rangle \right\vert
^{2}e^{-(2r+1)/\xi _{\perp }}\left( 1+\frac{\delta }{\xi _{\perp }}\right)
\label{xx2}
\end{equation}

Let us consider an alternation correlation functions \cite{Herzog}:%
\begin{equation}
\Delta _{\perp }(r)=\left\vert \left\langle
S_{n}^{x}S_{n+r}^{x}\right\rangle -\left\langle
S_{n}^{x}S_{n-r}^{x}\right\rangle \right\vert
\end{equation}

It equals zero for even $r$. But for large odd $r$ it becomes%
\begin{equation}
\Delta _{\perp }=s^{2}\left\vert \left\langle \psi _{00}|n_{x}|\psi
_{01}\right\rangle \right\vert ^{2}\frac{2\left\vert \delta \right\vert }{%
\xi _{\perp }}e^{-r/\xi _{\perp }}
\end{equation}

For the small and large $\gamma $ we obtain%
\begin{eqnarray}
\Delta _{\perp } &=&\frac{2T\left\vert \delta \right\vert }{3J(1-\delta ^{2})%
}e^{-r/\xi _{\perp }},\qquad \gamma \ll 1  \nonumber \\
\Delta _{\perp } &=&\frac{2T\left\vert \delta \right\vert }{J(1-\delta ^{2})}%
e^{-r/\xi _{\perp }},\qquad \gamma \gg 1  \label{D}
\end{eqnarray}

The comparison of correlation functions (\ref{Kg0}) and (\ref{D})
for $\gamma =0$ with those for the quantum dimerized FM model
obtained in \cite{Herzog} shows that they coincide in the leading
terms in $T$. Therefore, we claim that these correlation functions
for $r\gg 1$ obtained for the classical model are valid in scaling
limit for the quantum model as well.

Comparing Eqs.(\ref{xx1}) and (\ref{xx2}) one can see that the
corrections $\delta /\xi _{\perp }$ annihilate each other and
gives no contribution to the spin structure factor
\begin{equation}
S_{\bot }(q)=\frac{1}{N}\sum_{j,r}\left\langle
S_{j}^{x}S_{j+r}^{x}\right\rangle e^{iqr}
\end{equation}%
which is the sum of these equations. This is valid in the linear in
$\delta /\xi _{\perp }$ terms in the low-temperature limit. The
terms $\sim (\delta /\xi _{\perp })^{2}$ can introduce this
dimerization effect into $S_{\bot }(q)$, but this effect is out of
the scope of the used continuum approximation. Thus, in the leading
term in $\delta T/J$ the spin structure factor for the dimerized
model coincides with that of the FM model with the renormalized
exchange coupling. Using Eq.(\ref{nxx}) for the correlator
$\left\langle S_{j}^{x}S_{j+r}^{x}\right\rangle $ we obtain the
spin structure factor in the form%
\begin{equation}
S_{\bot }(q)=\frac{s^{4}J(1-\delta ^{2})}{T}\sum_{n}\frac{\left\vert
\left\langle \psi _{00}|n_{x}|\psi _{n1}\right\rangle \right\vert
^{2}(\varepsilon _{n1}-\varepsilon _{00})}{(\varepsilon _{n1}-\varepsilon
_{00})^{2}+\tilde{q}^{2}}
\end{equation}%
with $\tilde{q}=qs^{2}J_{0}/T$. The dependencies of the normalized
spin structure factor $\tilde{S}(q)=S_{\bot }(q)T/s^{4}J_{0}$ on
$\tilde{q}$ for several values of $\gamma $ is demonstrated in
Fig.3.

\begin{figure}[tbp]
\includegraphics[width=3in,angle=-90]{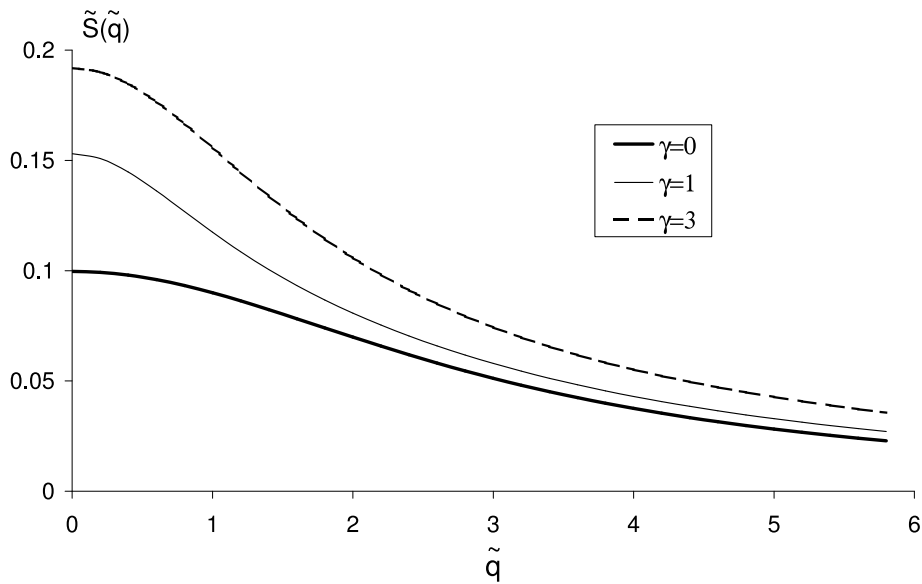}
\caption{Dependencies of the normalized spin structure factor
$\tilde{S}(q)=S_{\bot}(q)T/s^{4}J_{0}$ on the scaled wave vector
$\tilde{q}=qs^{2}J_{0}/T$ for $\gamma=0,1,3$.}
\end{figure}

\section{Spin-wave expansion of the quantum model}

According to the results of Sec.II the magnetization of the
classical dimerized model in the scaling limit $M(\gamma )$
coincides with that found in Ref.\cite{sachdev}. In
Ref.\cite{sachdev} a method of the computation of this function to
an arbitrary accuracy was developed. We cite several leading terms
of the expansion of $M(\gamma )$ for small and large values of
$\gamma $ obtained in Ref.\cite{sachdev}:
\begin{eqnarray}
M(\gamma ) &=&\frac{2}{3}\gamma -\frac{44}{135}\gamma ^{3}+\frac{752}{2835}%
\gamma ^{5}+...,\;\gamma \ll 1  \nonumber \\
M(\gamma ) &=&1-\frac{1}{2\gamma ^{1/2}}-\frac{1}{128\gamma ^{3/2}}-\frac{3}{%
512\gamma ^{2}}+.....,\;\gamma \gg 1  \label{Mg1}
\end{eqnarray}

However, it is not clear whether the function $M(\gamma )$ is
universal in the sense that it is valid for both the classical and
the quantum dimerized FM chains. Below we will produce arguments
in favor that such universality is the case.

At first, we compare the zero-field susceptibility given by%
\begin{equation}
\chi (0)=\frac{s^{4}J(1-\delta ^{2})}{T^{2}}\left. \frac{dM}{d\gamma }%
\right\vert _{\gamma =0}  \label{chi1}
\end{equation}%
with the asymptotic of $\chi (0)$ at $T\to 0$ obtained in Ref.\cite%
{Herzog} for the quantum dimerized FM chain. According to Ref.\cite{Herzog}%
\begin{equation}
\chi (0)=\frac{2}{3}\frac{s^{4}J(1-\delta ^{2})}{T^{2}}  \label{chi2}
\end{equation}

As follows from Eqs.(\ref{Mg1})-(\ref{chi2}) both expressions for
$\chi (0)$ coincide and function $M(\gamma )$ correctly describes
the limit $\gamma \to 0$ for the magnetization of the quantum model.
In connection with Eq.(\ref{chi1}) we note that it is not applicable
for the case of full dimerization, $\delta =1$. In this case the
system consists of decoupled dimers and the susceptibility follows
the Curie law $\chi (0)\sim 1/T$. Thus, our approach is valid when
$J(1-\delta ^{2})\gg T$.

Another check of the hypothesis of the universality is the comparison of the
spin-wave expansion for the quantum model with the expansion of $M(\gamma )$
for large $\gamma $ given by Eq.(\ref{Mg1}).

The spin-wave expansion is usually carried out by expressing the spin
operators using either the Holstein-Primakoff or the Dyson-Maleev
transformations. Here we use the latter which is%
\begin{eqnarray}
S_{n}^{+} &=&\sqrt{2s}(1-\frac{1}{2s}a_{n}^{+}a_{n})a_{n}  \nonumber \\
S_{n}^{-} &=&\sqrt{2s}a_{n}^{+}  \nonumber \\
S_{n}^{z} &=&s-a_{n}^{+}a_{n}  \label{DM}
\end{eqnarray}%
where $a_{n}^{+}$ and $a_{n}$ are the Bose-operators.

Using Eq.(\ref{DM}) we can write Hamiltonian (\ref{H3}) in terms of the
Bose-operators. The Bose analog of the spin Hamiltonian (\ref{H3}) contains
terms which are quadratic and quartic in the Bose-operators. The Fourier
transform to the momentum space operators leads to the Hamiltonian in the
form%
\begin{equation}
H=H_{0}+H_{int}-Ns^{2}J-Nsh  \label{Hsw}
\end{equation}%
where%
\begin{equation}
H_{0}=\sum_{k}[2sJ(1-\cos k)+h]a_{k}^{+}a_{k}+2i\delta
sJ\sum_{k}\sin ka_{k}^{+}a_{k+\pi }  \label{H0sw}
\end{equation}%
\begin{eqnarray}
H_{int} &=&\sum
V(k_{1}^{/},k_{2}^{/},k_{2},k_{1})a_{k_{1}^{/}}^{+}a_{k_{2}^{/}}^{+}a_{k_{2}}a_{k_{1}}\delta
(k_{1}^{/}+k_{2}^{/}-k_{2}-k_{1})
\nonumber \\
&&-i\delta \sum
W(k_{1}^{/},k_{2}^{/},k_{2},k_{1})a_{k_{1}^{/}}^{+}a_{k_{2}^{/}}^{+}a_{k_{2}}a_{k_{1}}\delta
(k_{1}^{/}+k_{2}^{/}-k_{2}-k_{1}-\pi ) \label{Hintsw}
\end{eqnarray}%
\begin{equation}
V(k_{1}^{/},k_{2}^{/},k_{2},k_{1})=-\frac{J}{4}\left[ \cos
(k_{1}-k_{1}^{/})+\cos (k_{2}-k_{1}^{/})+\cos (k_{1}-k_{2}^{/})+\cos
(k_{2}-k_{2}^{/})-2\cos (k_{1}^{/})-2\cos (k_{2}^{/})\right]  \label{V}
\end{equation}%
\begin{equation}
W(k_{1}^{/},k_{2}^{/},k_{2},k_{1})=-\frac{J}{4}\left[ \sin
(k_{1}-k_{1}^{/})+\sin (k_{2}-k_{1}^{/})+\sin (k_{1}-k_{2}^{/})+\sin
(k_{2}-k_{2}^{/})+2\sin (k_{1}^{/})+2\sin (k_{2}^{/})\right]  \label{W}
\end{equation}

Hamiltonian (\ref{H0sw}) can be diagonalized by a standard way. Let us
rewrite the Hamiltonian $H_{0}$ using the transformation of $k$-sums to
those over the reduced Brillouin zone and introducing the new Bose-operators
$\alpha _{k}$ and $\beta _{k}$ by the relation%
\begin{eqnarray}
a_{k}^{+} &=&u_{k}\alpha _{k}^{+}+v_{k}\beta _{k}^{+}  \nonumber \\
a_{k+\pi }^{+} &=&-iv_{k}\alpha _{k}^{+}+iu_{k}\beta _{k}^{+}  \label{cdab}
\end{eqnarray}%
where $\left\vert k\right\vert <\pi /2$ and%
\begin{eqnarray}
u_{k}^{2} &=&\frac{1}{2}+\frac{\cos k}{2\varepsilon (k)}  \nonumber \\
v_{k}^{2} &=&\frac{1}{2}-\frac{\cos k}{2\varepsilon (k)}  \nonumber \\
\varepsilon (k) &=&\sqrt{1-(1-\delta ^{2})\sin ^{2}k}  \label{uv}
\end{eqnarray}

Then, the Hamiltonian $H_{0}$ takes the form:%
\begin{equation}
H_{0}=\sum_{\left\vert k\right\vert <\pi /2}[E_{\alpha }(k)\alpha
_{k}^{+}\alpha _{k}+E_{\beta }(k)\beta _{k}^{+}\beta _{k}]
\label{H0albe}
\end{equation}%
where%
\begin{eqnarray}
E_{\alpha }(k) &=&2Js[1-\varepsilon (k)]+h  \nonumber \\
E_{\beta }(k) &=&2Js[1+\varepsilon (k)]+h  \label{Eab}
\end{eqnarray}

Hamiltonian (\ref{H0albe}) describes the non-interacting bosons. At $%
T\to 0$ the main contribution to the free energy from $H_{0}$ is
given by the small $k$ region. The expansion for $k\to 0$ results in%
\begin{eqnarray}
E_{\alpha }(k) &\simeq &sJ(1-\delta ^{2})k^{2}+h\;  \nonumber \\
\;E_{\beta }(k) &\simeq &4sJ  \label{exp}
\end{eqnarray}%
so that the thermal occupation numbers of $\alpha $ and $\beta $ particles
are%
\begin{eqnarray}
n_{\alpha }(k) &=&\frac{1}{e^{E_{\alpha }(k)/T}-1}\simeq \frac{T}{E_{\alpha
}(k)}  \nonumber \\
n_{\beta }(k) &\simeq &\exp \left( -\frac{4sJ}{T}\right) \to 0
\label{nab}
\end{eqnarray}

According to Eq.(\ref{nab}) we can omit in Eq.(\ref{H0albe}) the $\beta $
terms giving the exponentially small contribution to the thermodynamics at $%
T\to 0$. Then the Hamiltonian $H_{0}$ takes a form%
\begin{equation}
H_{0}=\sum [sJ(1-\delta ^{2})k^{2}+h]\alpha _{k}^{+}\alpha _{k}
\label{H02}
\end{equation}

Eq.(\ref{H02}) has a form of the Hamiltonian $H_{0}$ of the uniform FM model
with the renormalized exchange integral $J_{0}=J(1-\delta ^{2})$.

Now, let us consider the Hamiltonian $H_{int}$. First of all, we have to
express the operators $a_{k}$ in Eq.(\ref{Hintsw}) by the operators $\alpha
_{k}$ and $\beta _{k}$ using Eq.(\ref{cdab}). As was noted above, for
sufficiently low temperatures we can neglect the terms in $H_{int}$
containing $\beta _{k}^{+}$ and $\beta _{k}$ operators. Besides, we can
replace the Dyson-Maleev vertices $V$ and $W$ by their long-wavelength
limits. Carrying out some algebra for both terms in Eq.(\ref{Hintsw}) we
obtain $H_{int}$ in the form%
\begin{equation}
H_{int}=-\frac{1}{2}J(1-\delta ^{2})\sum k_{1}k_{2}\alpha
_{k_{1}^{/}}^{+}\alpha _{k_{2}^{/}}^{+}\alpha _{k_{2}}\alpha
_{k_{1}}\delta (k_{1}^{/}+k_{2}^{/}-k_{2}-k_{1})  \label{Hint2}
\end{equation}

A remarkable fact is that Eq.(\ref{Hint2}) is nothing but the quartic in the
Bose-operators part of the Hamiltonian of the uniform FM chain with the
renormalized exchange integral $J_{0}=J(1-\delta ^{2})$. In other words, the
Dyson-Maleev vertex of the dimerized chain is renormalized one of the
uniform model.

Thus, we established that in the long-wavelength limit, which is
justified at low temperatures, the Hamiltonian of the quantum
dimerized chain (\ref{Hsw}) reduces to that of the uniform FM
chain with the effective exchange integral $J_{0}=J(1-\delta
^{2})$. That is exactly as was found in Sec.II for the classical
spin chains. This fact implies that the known equivalence of the
low-temperature magnetic properties of classical and quantum FM
chains remains for the dimerized chains. It means that if the
universality relatively to the spin value holds for the uniform
model then this property remains valid for the dimerized FM chain
as well. Nevertheless, it is interesting to compare the large
$\gamma $ expansion for the quantum and the classical models.

The calculation of the spin wave expansion for the free energy and
the magnetization in all orders in $1/\gamma$ is a complicated
problem. But the leading terms can be found analytically. In the
zeroth order in $H_{int}$, which corresponds to the linear
spin-wave approximation, the magnetization
in the scaling limit is%
\begin{equation}
M^{(0)}=-\frac{1}{2\sqrt{\gamma }}
\end{equation}

It is easy to check that the contribution of the first order, $M^{(1)}$,
vanishes by a symmetry. The two-loop correction $M^{(2)}$ was calculated in
\cite{Kollar} and it is given by%
\begin{equation}
M^{(2)}=-\frac{1}{128\gamma ^{3/2}}
\end{equation}

Thus, the spin-wave expansion of the quantum dimerized FM chain is%
\begin{equation}
M=1-\frac{1}{2\sqrt{\gamma }}-\frac{1}{128\gamma ^{3/2}}+O(\gamma ^{-2})
\label{Mg2}
\end{equation}

The comparison of Eq.(\ref{Mg1}) with Eq.(\ref{Mg2}) shows that
both expansions are identical. Though we can not calculate the
spin-wave expansion in all orders, coincidence of the non-trivial
terms in Eq.(\ref{Mg2}) with those for $M(\gamma )$ is a strong
argument that the function $M(\gamma )$ gives the low-temperature
magnetization of both the classical and the quantum dimerized FM
chains.

\section{Discussion}

We studied the low-temperature magnetic properties of the
classical and quantum dimerized ferromagnetic spin chain. It is
shown that at low temperatures the classical dimerized model
reduces to the classical uniform model with the effective exchange
integral $J_{0}=J(1-\delta ^{2})$, where $\delta $ is the
dimerization parameter. The partition function and spin
correlation function of the classical model are calculated with
use of the mapping to the continuum limit, which is justified at
low temperatures. In the continuum limit the field-dependent
thermodynamics depends on one scaling parameter $\gamma
=hs^{3}J(1-\delta ^{2})/T^{2}$. The calculation of the partition
function and spin correlation function reduces to the solution of
the Schr\"{o}dinger equation for the quantum rotator in the
"gravitational" field $\gamma $.

We have studied the influence of the dimerization on the magnetic
properties of the classical spin model. In particular, we have
shown that the magnetization decreases with the increase of the
dimerization. We found the dependence of the spin correlation
functions on both the magnetic field and the dimerization
parameter. In contrast with the uniform model the correlation
functions as a function of the distance $r$ are different for the
even and odd $r$. Though the correlation lengths of the spin
correlations are the same for even and odd $r$, the preexponential
factors are different.

It was argued in Ref.\cite{sachdev} that the magnetization $M$ of
the uniform classical FM chain at $T\to 0$ is the universal
function of the scaling variable $g=s^{3}Jh/T^{2}$, i.e. the
universal function $M(g)$ is valid for any spin value $s$ and the
dependence on spin $s$ is captured in the scaling variable $g$
only. It implies that the magnetization curve $M(g)$ is valid for
both quantum and classical FM chains and can be determined by the
computation of the magnetization of the classical FM chain in the
so-called scaling limit, when $T\to 0$ and $h\to 0$ but the value
of $g$ is finite.

We have shown that this universality holds for the dimerized chain
as well. To confirm this fact we studied the quantum dimerized
spin model with the use of the Dyson-Maleev representation of spin
operators. It is shown that in the long-wavelength limit, which is
justified at low temperatures, the Hamiltonian of the quantum
dimerized chain reduces to that of the uniform quantum FM chain
with the effective exchange integral $J_{0}=J(1-\delta ^{2}) $.
That is exactly the same renormalization of the exchange coupling
as was found for the classical dimerized spin chains. This fact
implies that the known equivalence of the low-temperature magnetic
properties of classical and quantum FM chains remains valid for
the dimerized chains.

The physical reason of the equivalence of the low-temperature
magnetic properties of quantum and classical models is that the de
Broglie wavelength of spin waves $\lambda _{B}$ is less than the
ferromagnetic correlation length $\xi $ \cite{sachdev}. Indeed,
for the spectrum $Jk^{2}$ the de Broglie wavelength $\lambda
_{B}\sim 1/k\sim \sqrt{J/T\text{,}}$ while $\xi \sim J/T$. This
implies that the physical properties defined by the long-distance
asymptotics like magnetization are equal for quantum and classical
models. But on the short-distances the equivalence failed. For
example, the short-distance correlation function behaves as
$\left\langle S_{i}^{z}S_{i+r}^{z}\right\rangle \sim s^{2}-ar^{2}$
in quantum case \cite{Herzog}, while the classical model has
$\left\langle S_{i}^{z}S_{i+r}^{z}\right\rangle \sim s^{2}-br$.

The considered dimerized ferromagnetic model can be generalized by
including in Hamiltonian (\ref{H3}) the next--nearest-neighbor
antiferromagnetic exchange interaction
\begin{equation}
J_{13}\sum \mathbf{S}_{n}\cdot \mathbf{S}_{n+2},\quad J_{13}>0
\end{equation}

This term leads to the frustration. It is known \cite{exact} that the ground
state of the quantum dimerized chain with this interaction has the
ferromagnetic ground state for $\alpha =\frac{J_{13}}{J}<\frac{1-\delta ^{2}%
}{4}$ ($\alpha $ is the frustration parameter) and the singlet ground state
with the helical spin correlations for $\alpha >\frac{1-\delta ^{2}}{4}$.
For the classical model the transition from the ferromagnetic to the helical
phase occurs at the same value $\alpha =\frac{1-\delta ^{2}}{4}$. The
classical dimerized FM chain with the frustration can be studied in full
analogy with that for the pure ferromagnetic model. In particular, the
low-temperature thermodynamics is defined by the solution of the Schr\"{o}%
dinger equation (\ref{sch}) with the scaling parameter
\begin{equation}
\gamma =\frac{hs^{3}J(1-\delta ^{2}-4\alpha )}{T^{2}}
\end{equation}

It is easy to check that all presented results are valid for the frustrated
model. In particular, the free energy and the magnetization are equal to
those for the uniform FM chain with the renormalized exchange integral $%
J_{0}=J(1-\delta ^{2}-4\alpha )$. Therefore, we believe that the
magnetization is described by the universal function, which is valid for
both classical and quantum model if the frustration parameter $\alpha $ is
not too close to $\frac{1-\delta ^{2}}{4}$. The behavior of the
magnetization and the susceptibility in the the critical point $\alpha =%
\frac{1-\delta ^{2}}{4}$ it is radically different \cite{DK10,SKD}. For
example, the zero-field susceptibility $\chi (0)\sim T^{-4/3}$ in contrast
with $T^{-2}$ behavior for $\alpha <\frac{1-\delta ^{2}}{4}$.

\end{document}